\pdfoutput=1

\documentclass[11pt]{article}

\usepackage[final]{acl}
\usepackage{pifont}
\usepackage{tabularx}
\usepackage{booktabs}
\usepackage{caption}
\usepackage{multirow}
\usepackage{amssymb}
\usepackage{times}
\usepackage{latexsym}

\usepackage[T1]{fontenc}

\usepackage[utf8]{inputenc}

\usepackage{microtype}

\usepackage{inconsolata}

\usepackage{graphicx}

\usepackage{xspace}

\newcommand{\ttt}[1]{\texttt{#1}}
\newcommand{\tbf}[1]{\textbf{#1}}
\newcommand{\tss}[1]{\textsuperscript{#1}}
\newcommand{\tn}[1]{\textnormal{#1}}

\usepackage{enumitem}
\setlist[enumerate]{
    topsep=0pt,itemsep=-0.5ex,
    partopsep=0.5ex,parsep=0.5ex
}
\setlist[itemize]{
    topsep=0pt,itemsep=-0.5ex,
    partopsep=0.5ex,parsep=0.5ex,
}

\usepackage{booktabs}
\usepackage{colortbl}

\usepackage{mdframed}
\usepackage{listings}
\lstdefinestyle{minted}{
    basicstyle=\footnotesize\ttfamily\linespread{4},
    breaklines=true,
    columns=flexible,
    commentstyle=\color[rgb]{0.127,0.427,0.514}\ttfamily\itshape,
    identifierstyle=\color{black},
    inputencoding=latin1,
    keywordstyle=\color[HTML]{228B22}\bfseries,
    ndkeywordstyle=\color[HTML]{228B22}\bfseries,
    prebreak=\raisebox{0ex}[0ex][0ex]{\ensuremath{\hookleftarrow}},
    showstringspaces=true,
    stringstyle=\color[rgb]{0.639,0.082,0.082}\ttfamily,
    upquote=true
}
\surroundwithmdframed[
    backgroundcolor=gray!10, linecolor=gray!60,
    innertopmargin=0, innerbottommargin=0
]{lstlisting}

\usepackage{fontawesome5}

\usepackage{todonotes}

\makeatletter
\newcommand\equalcontrib[1]{%
  \begingroup
  \let\thefootnote\relax
  \footnote{#1}%
  \addtocounter{footnote}{-1}%
  \endgroup
}
\makeatother

\title{Synthetic Audio Helps for Cognitive State Tasks}

\author{
    Adil Soubki\tss{\tn{$\ast$$\blacklozenge\spadesuit$}},
    John Murzaku\tss{\tn{$\ast$$\blacklozenge\spadesuit$}},
    Peter Zeng\tss{\tn{$\blacklozenge\spadesuit$}},
    Owen Rambow\tss{\tn{$\clubsuit\spadesuit$}} \\
    \tss{$\blacklozenge$}Department of Computer Science
    \tss{$\clubsuit$}Department of Linguistics\\
    \tss{$\spadesuit$}Institute for Advanced Computational Science\\
    Stony Brook University\\
    \texttt{\{asoubki,jmurzaku\}@cs.stonybrook.edu}
}

\begin{document}
\maketitle
\begin{abstract}
Automatically recognizing a human's complete cognitive state from text is a difficult task; from text, a model has to recognize a combination of concepts including belief, emotion, common ground, sentiment, and intention. 
Humans do not only track and update cognitive state from the meaning of words and sentences, but also from paralinguistic cues such as prosody.
The NLP community has broadly focused on text-only approaches to cognitive state tasks, but audio can provide vital missing information. We posit that text-to-speech (TTS) models learn to track aspects of cognitive state in order to produce naturalistic audio, and that the signal audio models implicitly identify is orthogonal to the information that language models exploit. We present Synthetic Audio Data fine-tuning (SAD), a framework where we show that seven tasks related to cognitive state modeling benefit from multimodal training on both text and zero-shot synthetic audio data from an off-the-shelf TTS system. We show an improvement over the text-only modality when adding synthetic audio data to text-only corpora. Furthermore, on tasks and corpora that do contain gold audio, we show our SAD framework achieves competitive performance using text and synthetic audio compared to text and gold audio.
\end{abstract}

\section{Introduction}
A significant amount of work in NLP focuses on tasks that involve extracting information about the cognitive states of human entities from text. This includes predicting beliefs (or ``event factuality'') \citep{sauri2009factbank}, recognizing emotions \citep{canales-martinez-barco-2014-emotion}, recognizing sentiment \citep{wiebe-1990-identifying}, tracking common ground \citep{markowska-etal-2023-finding}, predicting conversation success \citep{zhang-etal-2018-conversations}, identifying intentions \citep{colombo2020guiding}, among others. \equalcontrib{$^\ast$Denotes equal contribution}

Prior work has shown that audio signals, when available, improve performance for a number of tasks involving cognitive states \citep{murzaku24_interspeech, zhao22k_interspeech, nojavanasghari2016deep}. Meanwhile, text-to-speech (TTS) systems have improved rapidly over the past several years -- particularly when it comes to synthesizing more naturalistic audio. Part of what models must learn in order to generate realistic speech involves saying words in a way that matches the cognitive states those words reveal. In this paper we investigate two research questions related to these observations.

\textbf{RQ1:} In the event that one has a task with human audio, how does using synthetic (TTS generated) audio compare to the gold-standard (human) audio?
\textbf{RQ2:} Can synthetic audio help even for tasks which never had human audio to begin with?

We hypothesize that synthetic audio will perform worse than human audio (\tbf{RQ1})  but better than text-only (no audio). Furthermore, the aspects of cognitive state that TTS models learn to predict in order to produce naturalistic speech will provide orthogonal signal to the patterns text-based language models pick up on and improve performance, even on datasets for which human audio is not available (\tbf{RQ2}). %

Our main contribution is to present SAD, a multimodal synthetic audio data framework that boosts performance on cognitive state tasks that do not contain audio, or offers competitive performance on tasks that do.

The paper is organized as follows. A survey of previous work is provided in Section~\ref{sec:prev}. We summarize the SAD framework in Section~\ref{sec:sad} and present our experiments on SAD in Section~\ref{sec:expt}. We conclude and provide a discussion and implications of our novel framework in Section~\ref{sec:concl}.

We emphasize that this paper does not introduce new machine learning architectures; instead, we show that synthetic audio data through our SAD framework can lead to improved performance without the necessity of introducing new, more complex architectures. We view our framework as a generalizable solution; it can profit and adapt from advances in language models, TTS models, and multimodal models. We release our framework and models on GitHub\footnote{\url{https://github.com/adil-soubki/sad-training}}.

\section{Related Work}
\label{sec:prev}
To the best of our knowledge, we are the first to present a multimodal text and audio framework with synthetic audio data from TTS systems for cognitive state tasks.
However, regarding experiments with audio signal, there has been previous work on multimodal (text and audio) and unimodal (audio only) models for corpora in emotion, belief, deception, and sentiment. 

\paragraph{Multimodal} There has been some work in fusing text and audio features for cognitive state tasks, specifically in emotion and belief. In emotion, 
\citet{zhao24g_interspeech} present a novel architecture containing a refined attention mechanism, a novel perception unit aligning the emotion frame to the global audio context, and a new convolution procedure to effectively fuse audio and text features.
\citet{kyung24_interspeech} fine-tune BERT \citep{devlin-etal-2019-bert} and fuse with ASR features derived from speech, achieving state-of-the-art results on the multimodal emotion corpus IEMOCAP \citep{busso2008iemocap}, which we also test our SAD framework on.
In the multimodal belief prediction task, \citet{murzaku24_interspeech} were the first to show that fusing text with audio features helps, achieving state-of-the-art results on the CB-Prosody corpus \citep{mahler2020prosody}. 

Regarding deception, there has been previous work on acoustic and lexical approaches.  Testing on the CXD corpus \citep{levitan2015cross}, \citet{Mendels2017HybridAD} show that a multimodal architecture boosts performance compared to a unimodal text-only approach.

\paragraph{Audio Only} There has also been work focusing on the audio-only modality for cognitive state tasks, but considerably less than multimodal. For the deception detection task, \cite{levitan18_interspeech,chen2020acoustic,levitan2022believe} focus on training classical machine learning methods with acoustic and prosodic features. Regarding emotion detection, \citet{pepino2021emotion} were the first to fine-tune a pre-trained speech model for emotion detection, specifically Wav2Vec2.0 \citep{baevski2020wav2vec}.

\section{SAD Overview}
\label{sec:sad}
\begin{figure}
    \centering
    \includegraphics[width=\linewidth]{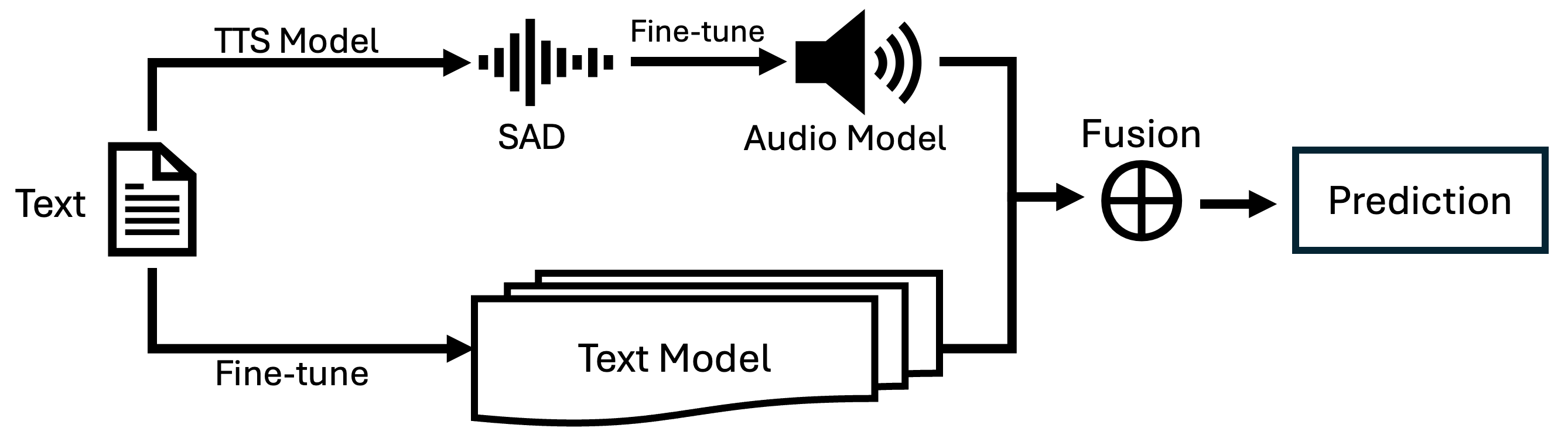}
    \caption{Overview of the SAD framework, beginning with a text input. We then perform zero-shot TTS on the text to get audio and then fine-tune an audio model. In parallel, we fine-tune a text model. We then fuse the features from both modalities to get a final prediction.}
    \label{fig:sad-diagram}
\end{figure}

\paragraph{Text} We perform all fine-tuning experiments with BERT, specifically bert-base-uncased. We closely mirror the experimental setup of previous work on text and audio multimodal cognitive state architectures which find BERT to be the top performing text encoder \citep{zhao22k_interspeech,murzaku24_interspeech}.\looseness=-1

\paragraph{TTS: OpenAI} We generate synthetic audio data using the OpenAI TTS API \citep{openai_tts}. We specifically generate all of our data using the Alloy voice and the tts-1-hd model which is optimized for audio quality. While we also performed experiments with different voices, we fix to one voice (Alloy) due to API costs. 

\paragraph{TTS: Open Source} To emphasize the generalizability and viability of our SAD framework, we also perform experiments with an open source TTS model, specifically MatchaTTS \citep{mehta2024matcha}. We choose MatchaTTS due to its minimal memory requirements, competitive speed on long utterances, and top performance in terms of mean opinion score in human evaluations.
We also experimented with other TTS APIs such as \citep{elevenlabs}, but chose to stick with OpenAI API and Matcha as they were the most cost effective and yielded the highest quality audio.

\paragraph{Audio} We follow \citet{murzaku24_interspeech} and use the pre-trained Whisper model \cite{radford2023robust} as our audio encoder, specifically whisper-base. For all Whisper experiments, we pad all audio clips to the maximum 30 second limit. 

\paragraph{Multimodal} We combine all the previously listed individual components into a unified multimodal architecture. We show this architecture in Figure~\ref{fig:sad-diagram}. We perform both early fusion and late fusion experiments, similar to \citep{murzaku24_interspeech,zhao22k_interspeech,nojavanasghari2016deep}
: in our late fusion model, BERT and Whisper are fine-tuned separately with their representations max pooled and concatenated after passing through individual regression heads. In the early fusion model, the final hidden representations are max pooled and concatenated before being jointly fine-tuned and passed through a shared regression head.
We emphasize the generalizability of the SAD framework: each individual component can be replaced with fine-tuned or task-specific models.\looseness=-1

\section{Experiments}
\label{sec:expt}
\begin{table*}[t]
\setlength\tabcolsep{4.1pt}
\centering
\small
\begin{tabular}{llccclclcccccc}
\toprule
\multirow{2.5}{*}{\textbf{Type}} & \multirow{2.5}{*}{\textbf{Task}} & \multicolumn{2}{c}{\textbf{Gold}} & \multirow{2.5}{*}{\textbf{Size}} & \multirow{2.5}{*}{\textbf{Metric}} & \multirow{2.5}{*}{\textbf{Text}} & \multicolumn{3}{c}{\textbf{Audio}} & \multicolumn{3}{c}{\textbf{Multimodal}} \\
\cmidrule(lr){3-4} \cmidrule(lr){8-10} \cmidrule(lr){11-13}
 &  & Text & Audio &  &  &  & Gold & Matcha & OpenAI & Gold\tss{$\dagger$} & Matcha & OpenAI\tss{$\dagger$} \\
\midrule
\multirow{3}{*}{Control} & BoolQ {\tiny\faSeedling\hspace{0.25em}\faCut} & \ding{51} & \ding{55} & 509\tss{\tiny\faAngleDoubleDown} & Acc $\uparrow$ & 60.0 & - & \tbf{69.7} & 69.4 & - & 65.3 & 67.1 \\
 & WIC {\tiny\faSeedling} & \ding{51} & \ding{55} & 6,066 & Acc $\uparrow$ & \tbf{57.3} & - & 49.1 & 47.4 & - & 59.6 & 57.1 \\
 & WSC {\tiny\faSeedling} & \ding{51} & \ding{55} & 658 & Acc $\uparrow$ & \tbf{63.5} & - & \tbf{63.5} & 60.6 & - & \tbf{63.5} & \tbf{63.5} \\
\midrule
\multirow{2}{*}{Sentiment} & SWBD-S {\tiny\faFolderOpen} & \ding{51} & \ding{51} & 2,856\tss{\tiny\faAngleDoubleDown} & MAE $\downarrow$ & 0.339 & 0.465 & 0.461 & 0.462 & \tbf{0.334} & 0.341 & \tbf{0.334} \\
 & IMDB {\tiny\faSeedling\hspace{0.25em}\faCut} & \ding{51} & \ding{55} & 372\tss{\tiny\faAngleDoubleDown} & Acc $\uparrow$ & 89.5 & - & 63.5 & 58.5 & - & 88.3 & \tbf{89.7} \\
\midrule
\multirow{3}{*}{Belief} & CB-Prosody {\tiny\faFolderOpen} & \ding{51} & \ding{51} & 334 & MAE $\downarrow$ & 0.693 & 1.083 & 0.931 & 0.906 & \tbf{0.665} & 0.699 & 0.668 \\
 & CB {\tiny\faSeedling} & \ding{51} & \ding{55} & 500 & MAE $\downarrow$ & 0.785 & - & 1.189 & 1.154 & - & 0.756 & \tbf{0.741} \\
 & FactBank {\tiny\faSeedling} & \ding{51} & \ding{55} & 7,540 & F1 $\uparrow$ & 74.9 & - & 68.7 & 66.3 & - & \tbf{76.0} & \tbf{76.0} \\
\midrule
\multirow{2}{*}{Emotion} & IEMOCAP {\tiny\faSeedling} & \ding{51} & \ding{51} & 7,529 & F1 $\uparrow$ & 56.6 & 55.5 & 51.7 & 52.2 & \tbf{63.4} & 57.6 & 59.3 \\
 & GoEmotions {\tiny\faSeedling} & \ding{51} & \ding{55} & 4,753\tss{\tiny\faAngleDoubleDown} & F1 $\uparrow$ & 51.4 & - & 38.3 & 36.2 & - & 52.7 & \tbf{53.1} \\
\bottomrule
\end{tabular}
\caption{Overview of cognitive state tasks, including gold data availability, corpus size, evaluation metrics, and results for various modalities. Size represents the number of samples in the dataset ({\tiny\faAngleDoubleDown}\ indicates the dataset was down-sampled due to cost). We report Acc $\uparrow$ (Accuracy $\uparrow$), F1 $\uparrow$ (F1 $\uparrow$ score), and MAE $\downarrow$ (Mean Absolute Error) as metrics depending on the corpus. If the dataset did not have a canonical split, metrics are averaged over five folds (indicated by {\tiny\faFolderOpen}). Otherwise the metrics are averaged over three seeds (indicated by {\tiny\faSeedling}). If the audio data exceeded Whisper's 30 second context, it was truncated when training (datasets significantly affected are indicated by {\tiny\faCut}).
A binomial test ($\pi_0 = 0.5$) is used to determine if the frequency models outperform the text-only baseline on non-control tasks is significant ($\dagger$ indicates $p < 0.05$).
}
\label{tab:results}
\end{table*}

\subsection{Tasks}
A broad overview of the tasks
we test SAD on is shown in Table~\ref{tab:results}. We specifically test on four types of tasks: control tasks which are not about the writer's cognitive state, and for which we hypothesize that synthetic audio data will not improve compared to text only; sentiment; belief; and emotion. We also show the included modalities from each corpus; only three corpora contain both text and audio (SWBD-S, CB-Prosody, and IEMOCAP). We describe each task and its corpora in detail.

\paragraph{Control Tasks} Our control tasks include three tasks chosen from SuperGLUE \cite{wang-etal-2019-superglue}: BoolQ \citep{clark-etal-2019-boolq} (question answering), WiC \cite{pilehvar-camacho-collados-2019-wic} (word sense disambiguation),
and WSC \citep{winograd2012} (pronoun resolution and common sense reasoning). We choose these three tasks as they do not explicitly model cognitive state, expecting SAD to not improve performance.

\paragraph{Sentiment} We test on two corpora that annotate for sentiment. The Switchboard Sentiment (SWBD-S) corpus \citep{chen-etal-2020-large} annotates segments of Switchboard \citep{godfrey1992switchboard} for sentiment averaged among three annotators, resulting in a continuous sentiment value from [-1, 1]. This corpus contains gold audio; we therefore compare our proposed SAD framework to multimodal experiments with the gold audio.
We also test on the IMBD corpus \citep{maas-etal-2011-learning} which is a standard benchmark for author sentiment in NLP.

\paragraph{Belief} The term ``belief'' refers to how committed is the author or speaker to the truth of a proposition. \citet{murzaku24_interspeech} were the first to show that multimodal architectures with specifically text and speech signal boost performance on belief tasks compared to standard text only approaches. We therefore test whether SAD can help on three belief corpora: CB-Prosody \citep{mahler2020prosody} which \citet{murzaku24_interspeech} tested on, CB \citep{de2019commitmentbank}, and FactBank \citep{sauri2009factbank}. In CB, annotators were given transcripts of speaker utterances or written text
and asked to evaluate the level of certainty (or belief) the speaker appears to have regarding the truth of the proposition, specifically the content of the complement clause. Expanding on CB, CB-Prosody, which contains gold audio data, annotates 350 Switchboard examples present in CB, but annotators {\em only} heard the audio clips rather than reading transcripts. Both of these corpora have continuous annotations; specifically, belief values fall from a continuous range of [-3,3].\looseness=-1

We also choose FactBank \citep{sauri2009factbank} as a benchmark corpus for SAD, as it is one of the first carefully constructed datasets for belief prediction. We specifically use the author only examples and the split from \citet{murzaku-etal-2022-examining}. FactBank uses categorial labels.

\paragraph{Emotion}
We include two corpora for emotion recognition, since emotion is frequently expressed in audio. In order to synthesize audio which is naturalistic, TTS systems must learn to recognize utterances which reveal emotions, especially those of higher intensity, and vary their output accordingly. As such we expect there to be some signal from TTS for emotion recognition both when the dataset contains gold audio (IEMOCAP, \citep{busso2008iemocap}) and when it does not (GoEmotions, \citep{demszky2020goemotions}).

\subsection{Experimental Setup}

\paragraph{Training} For all experiments, we add either a classification or regression head depending on the task. All experiments are trained for 10 epochs; no hyperparameter tuning is performed.
We use a learning rate of 2e-5 and batch size of one. 

\paragraph{Data} To save API costs, we randomly downsample datasets until the cost of generating audio using OpenAI's TTS system costs \$10 USD per corpus. We 
use the same subcorpus when doing comparison experiments with Matcha-TTS.
We mark the datasets that were downsampled in Table~\ref{tab:results}. 
We provide more details on the data and experiments in Appendix~\ref{app:b}.

We note that we experimented with better or larger text models, specifically RoBERTa \citep{liu2019roberta} and Flan-T5 \citep{chung2024scaling}, but did not notice an improvement in our multimodal results over BERT as a text encoder (which is in line with findings by \citet{murzaku24_interspeech}).

\subsection{Evaluation} Table~\ref{tab:results} shows the metrics we evaluate each corpus on. We perform regression on three corpora, since they contain continuous values as annotations: SWBD-S, CB-Prosody, and CB. We therefore evaluate these three corpora on mean absolute error (MAE). For the rest of the corpora, which contain categorical labels, we perform an accuracy or F-measure evaluation. We indicate the metric for each corpus in the table, and whether more or less is better.
If the TTS audio improves model performance, we would expect models which use that signal to outperform those which do not more frequently than random across the non-control datasets.  We therefore use a binomial test ($\pi_0 = 0.5$) to determine the significance of this frequency. Where relevant, we also note differences within tasks using independent paired $t$-tests.\looseness=-1

\subsection{Results}
We first examine each set of experiments by task category and then conclude with a discussion of 
our research questions.

\paragraph{Control}
Since we hypothesize that the additional signal that TTS models provide comes from their ability to represent cognitive states through the way they synthesize audio, the control tasks were chosen 
as tasks which do not
reflect the cognitive state of the speaker. 
We expect multimodal training to not affect the results here and this is indeed the result for WIC and WSC. However, BoolQ performs best when given \emph{only} the audio data, which is surprising. This is likely due to BERT's poor handling of such long sequences of context before the question.
Since the audio model can only handle 30 seconds of context, we truncate the remaining audio
which seems to have helped for some reason.
We expect that further tuning
of the input representations would bring this result in line with the other control tasks.\looseness=-1

\paragraph{Sentiment}
Since sentiment analysis requires predicting the author's feelings (either in general or towards a particular aspect), we expect audio to improve our results. After averaging across all folds and seeds we do see this but the effect size is somewhat smaller than expected. For SWBD-S, the multimodal models just barely edge out the text-only variant, with gold audio and OpenAI synthetic audio both achieving the best MAE of 0.334. The SAD model using OpenAI TTS similarly performs best by a narrow margin on IMDB, scoring an accuracy of 89.7\%. As the gold audio in the SWBD-S corpus doesn't help much for sentiment, 
it makes sense that synthetic audio does not help either: apparently sentiment is mainly conveyed lexically in our sentiment corpora.

\paragraph{Belief}
The belief expressed by an utterance can entirely change based on the way the utterance is said (e.g., \emph{*John* said it's true} vs. \emph{John said, it's *true*}). 
It is possible that TTS systems have useful priors in such cases. 
Across all three belief datasets, after averaging metrics over all folds and seeds, the best performing models are the multimodal variants. For CB-Prosody, the gold audio results in a 4\% decrease in MAE and the OpenAI TTS model trails closely behind with a 3.6\% reduction in MAE. For the text-only CB, both Matcha and OpenAI synthetic audio models improve over the text-only baseline to achieve a 3.7\% and 5.6\% decrease in MAE, respectively. FactBank also sees a 4.4\% F1 error decrease
over the text-only model for both Matcha and OpenAI SAD models. 

\paragraph{Emotion}
In order to synthesize audio which is naturalistic, TTS systems must learn to recognize utterances which reveal emotions, especially those of higher intensity, and vary their output accordingly. As such we expect there to be some signal from TTS for emotion recognition in both the condition where the dataset contains gold audio (IEMOCAP) and when it does not (GoEmotions). 
For IEMOCAP we see a 6.8 point improvement in F1 over the text-only model when using gold audio. The synthetic audio models also outperform text-only with OpenAI and Matcha seeing a 1.0 and 1.7 point improvement, respectively. For GoEmotions, improvements of similar scale over the text-only are observed with Matcha showing a 1.3 point boost and OpenAI showing a 1.7 point boost. These differences are significant ($p < 0.05$) using independent two-sample $t$-tests.

\paragraph{Matcha vs. OpenAI} Though the audio-only models typically performed far worse than the multimodal variants, one might expect the TTS system which performed the best on audio-only fine-tuning (typically, Matcha) to perform the best in SAD fine-tuning. Our results do not support such a trend. TTS generations from OpenAI matched or outperformed Matcha in multimodal fine-tuning for all seven non-control tasks, though often by a small margin. In both cases early fusion tended to perform best.\looseness=-1  %

\paragraph{RQ1: How does SAD compare to using gold audio?} Generally, our experiments found SAD to perform worse than using gold audio but better than no audio at all. For SWBD-S and CB-Prosody, SAD (using OpenAI TTS) matched or nearly matched the performance of gold audio; for IEMOCAP, there was a sizable degradation (4.1 points F1) between the best SAD model and gold performance as discussed above.
If we consider just the noncontrol datasets with gold audio, a binomial test (using all experiments with either multiple seeds or folds, as shown in Table~\ref{tab:results}) indicates that the frequency with which the OpenAI multimodal models outperform their text-only counterparts is significant ($p < 0.05$). This is not the case for Matcha. 
The same analysis across gold and SAD models also indicates gold models perform better than SAD models significantly ($p < 0.05$) often.
When comparing performance of fusion models which receive gold audio against those that receive synthetic audio, IEMOCAP is the only case where a significant difference ($p < 0.05$) is observed using an independent two-sample $t$-test. This is also the only case where gold audio significantly ($p < 0.05$) improves the unimodal models. In other words, emotion tasks were particularly sensitive to audio quality.\looseness=-1

\paragraph{RQ2: Does SAD help data with no gold audio?} Yes.
Our experiments show that datasets that never had audio to begin with also see an improvement in performance. %
If we consider just the non-control datasets, binomial testing finds OpenAI (but not Matcha) audio to improve performance significantly ($p < 0.05$) often, similar to the case for datasets with gold audio.
Performing the same analysis for the control datasets, as expected, does not find significance.
The absence of (unexplained) improvements in the control tasks, suggests that TTS models contain latent signals for cognitive states.

\section{Conclusion}
\label{sec:concl}
We have introduced a new approach to prediction tasks about the cognitive state of a speaker or writer.  We show that using TTS to create synthetic audio helps across seven tasks when used in conjunction with text, compared to using only text.
While the effect sizes are currently small, performance gains will likely grow as TTS systems improve over the coming years.
Our research suggests that exploiting additional modalities, even when synthetic, may be a useful strategy in NLP tasks if we have reason to believe that the additional modality may carry orthogonal signal for the task.\looseness=-1

\section*{Limitations} 
\paragraph{Cognitive State Focus} Our work focuses on a targeted subset of tasks within cognitive state modeling. We understand that, while our SAD framework supports our hypothesis that TTS models capture cognitive state features, the idea may not be generalizable to broader NLP tasks. We leave this to future work and intend to explore broader tasks.
\paragraph{TTS Model Choice} A large portion of SAD primarily focuses on using closed API TTS models. We therefore understand that some areas may lack model details and implementation details. We however will release all scripts and details for generating our data through the API.

\section*{Ethical Considerations}

As with other work on cognitive states, we risk the misinterpretation that AI models may be anthromorphized
as having near-human level cognition. We stress
that our work shows that our framework can help text-only models when presented with natural sounding audio, but does not give AI models full cognitive state understanding or cognition.

We note that our paper is foundational research and we are not tied to any direct applications. 

\section*{Acknowledgements}
This material is based upon work supported in part by the National Science Foundation (NSF) under No. 2125295 (NRT-HDR: Detecting and Addressing Bias in Data, Humans, and Institutions); by funding from the Defense Advanced Research Projects Agency (DARPA) under the CCU project (No. HR001120C0037, PR No. HR0011154158, No. HR001122C0034); as well as by the Intelligence Advanced Research Projects Activity (IARPA) under the HIATUS program (contract 2022-22072200005).  Any opinions, findings and conclusions or recommendations expressed in this material are those of the author(s) and do not necessarily reflect the views of the NSF, DARPA, or IARPA.\looseness=-1

We thank both the Institute for Advanced Computational Science (IACS) and the Institute for AI-Driven Discovery and Innovation at Stony Brook for access to the computing resources and API fees needed for this work. These resources were made possible by NSF grant No. 1531492 (SeaWulf HPC cluster maintained by Research Computing and Cyberinfrastructure) and NSF grant No. 1919752 (Major Research Infrastructure program), respectively.\looseness=-1

We thank our ARR reviewers, whose comments have contributed to improving the paper.
\bibliography{anthology, custom}

\appendix
\section{Experiment Details}
All experiments besides our OpenAI experiments used our employer’s GPU cluster. We performed experiments on a Tesla V100-SXM2 GPU. Compute jobs typically ranged from 5 minutes for zero-shot TTS generation to 6 hours for multimodal fine-tuning. %
The text model used was \ttt{google-bert/bert-base-uncased} (110M params) and the audio model used was \ttt{openai/whisper-base}  (72.6M params).
We fine-tune all models for a fixed 10 epochs and report the relevant metrics at the last epoch. All experiments use a learning rate of 2e-5 and a batch size of 1. We do not perform any hyperparameter tuning or hyperparameter searches. We use the mean squared error (MSE) loss function for regression tasks and cross-entropy loss for classification tasks. We pad text to BERT's maximum sequence length of 512 and audio clips to Whisper's maximum sequence length of 30 seconds. We checked training loss curves to ensure that models were converging.

We report the average over three seeds (42, 0, 1) for corpora with an established train/test/dev split. For other corpora, we perform five-fold cross-validation and report the average over all five folds.

\section{Data Processing}
\label{app:b}
\textbf{Synthetic Audio Data} Prior to generating synthetic audio, for text with context longer than the target text for which a task is to be performed on, we shorten the text to just the target span. This is because prosody of generated audio tends to degrade as the lengths of the generations increase. To save costs, we down-sample datasets such that the cost of generating audio using OpenAI's TTS system costs \$10 USD, and match that data when generating audios with Matcha-TTS.

\textbf{FactBank} For datasets such as FactBank \citep{sauri2009factbank}, which annotate single event tokens, we extract the syntactic span using spaCy \citep{spacy}. We create a custom \ttt{head2span} module which takes the event head word and returns the syntactic span.

\textbf{CB} The CB dataset \citep{de2019commitmentbank} contains three sentences: two previous sentences of context, and the target sentence where the matrix clause is annotated. For all experiments, we use only the last, or the target sentence.

\textbf{SWBD-S} The SWBD-S corpus contains three annotations for sentiment averaged among three annotators resulting in a continuous sentiment value from [-1, 1]. We manually process this using our own scripts.

\end{document}